\documentclass[twocolumn,showpacs,prb]{revtex4}% Physical Review B

\usepackage{graphicx}%
\usepackage{dcolumn}
\usepackage{amsmath}

\begin{document}

\preprint{HEP/123-qed}

\title{ Curie temperature enhancement of electron doped Sr$_2$FeMoO$_6$ perovskites studied by photoemission spectroscopy }

\author{J. Navarro and J. Fontcuberta}

\address{Institut de Ci\`{e}ncia de Materials de Barcelona Campus U.A.B.,
Bellaterra 08193, Catalunya, Spain}

\author{M. Izquierdo, J. Avila and M. C. Asensio}

\address{LURE, Centre Universitaire Paris-Sud, B\^at. 209D,
B.P. 34, 91405 Orsay Cedex, France}
\address{Instituto de Ciencia de Materiales,
CSIC, 28049 Madrid, Spain}

\date{\today}

\begin{abstract}
We report here on the electronic structure of electron-doped
half-metallic ferromagnetic perovskites such
Sr$_{2-x}$La$_x$FeMoO$_6$ ($x$=0-0.6) as obtained from
high-resolved valence-band photoemission spectroscopy (PES). By
comparing the PES spectra with band structure calculations, a
distinctive peak at the Fermi level (E$_F$) with predominantly
(Fe+Mo) t$_{2g}^\downarrow$ character has been evidenced for all
samples, irrespectively of the $x$ values investigated. Moreover,
we show that the electron doping due to the La substitution
provides selectively delocalized carriers to the
t$_{2g}^\downarrow$ metallic spin channel. Consequently, a gradual
rising of the density of states at the E$_F$ has been observed as
a function of the La doping. By changing the incoming photon
energy we have shown that electron doping mainly rises the density
of states of Mo parentage. These findings provide fundamental
clues for understanding the origin of ferromagnetism in these
oxides and shall be of relevance for tailoring oxides having still
higher T$_C$.

\end{abstract}


\maketitle

%\tableofcontents

\section{Introduction}

The early discovery of colossal magnetoresistance (CMR) has
abundantly stimulated the research on manganese oxides because of
their leading technologic applications, not counting their
critical importance in basic physics. However, typically the
magnetoresistance strength diminishes as T$_C$ increases up to
room temperature, thus making it difficult the direct technologic
use of these advanced materials. As progress of spintronics
requires fully spin-polarized ferromagnetic materials, having
Curie temperatures (T$_C$) well above room temperature, recent
finding of room-temperature tunneling magnetoresistance and
half-metallic (HM) behavior of Sr$_2$FeMoO$_6$ (SFMO) oxides with
T$_C\sim$400K \cite{Kobnat} has opened renewed interest and
expectative for promising applications. A$_2$MM'O$_6$ double
perovskites are built up by alternate M'O$_6$ and MO$_6$
octahedral units bonded by oxygen bridges, where A is an alkaline
earth or rare-earth ion and M, M' are 3d and 4d/5d transition
metals. These novel materials are predicted to be HM
\cite{Kobnat,Kobprb,Pickett,Wu}, where the metallic behavior of
one electron spin channel is coexisting with an energy gap between
valence and conduction bands for electrons of the other spin
polarization. In short, the electronic configuration of SFMO can
be described by Fe(3d$^{6-\delta}$):Mo(4d$^\delta$) ($\delta$=0.3
\cite{Balc}). The Fe-3d full-filled (t$_{2g}^3$ and e$_g^2$)
$\it{spin}$-$\it{up}$ electrons can be viewed as localized whereas
the Fe 3d partially empty (t$_{2g}$$^{1-\delta}$)
$\it{spin}$-$\it{down}$ states are strongly hybridized with O-2p
orbitals and partially empty Mo-4d$^\delta$
($\it{spin}$-$\it{down}$) states. Consequently, these
transition-metal oxides alloys present bands associated to one
spin polarization (in this case Fe-3d $\it{spin}$-$\it{up}$)
entirely filled, which are separated from unoccupied bands of the
same symmetry by a bandgap sited at the Fermi level. Whereas, the
other spin channel involving Fe-3d and Mo-4d
$\it{spin}$-$\it{down}$ electrons has a metallic character.

Recent photoemission spectroscopy (PES) experiments
\cite{Sarprl,Kang,Saitoh} have provided evidence of the half
metallic character of SFMO by determining that the states close to
the Fermi level have predominantly Mo t$_{2g}^\downarrow$ and Fe
t$_{2g}^\downarrow$ character. By comparing photon-energy
dependence of the valence band spectra and the local spin-density
approximation (LSDA) band structure theoretical calculations has
revealed that the delocalized electrons at the E$_F$ are totally
$\it{spin}$-$\it{down}$ polarized. Consistent results have been
lately obtained by using X-ray absorption spectroscopy (XAS)
experiments \cite{Ray}. The site-specific reported information is
also in agreement with LSDA theoretical calculations that
contemplate well-localized t$_{2g}^3$ and e$_g^2$
$\it{spin}$-$\it{up}$ Fe-3d sub-bands are well below the Fermi
level and that the delocalized t$_{2g}$ $\it{spin}$-$\it{down}$
sub-bands due to the Mo(4d) and Fe (3d) are at the E$_F$. These
spectroscopic results together with recent magnetic measurements
in the paramagnetic phase \cite{Tovar} support that the AFM
interaction is driven by a mechanism where the itinerant carriers
and the Fe localized cores tend to be antiparallel, at variance
with the double exchange interaction.

Magnetic measurements are consistent with ferromagnetic ordering
of Fe(3d$^{6-\delta}$) moments which shall be
antiferromagnetically coupled to any moment on Mo(4d$^\delta$)
sites. Understanding of the physical mechanism lying behind the
ferromagnetic ordering remains challenging. Difficulties arise
mainly due to the fact that in this structure the 4d(Mo) ions are
essentially non-magnetic and thus the separation between the
magnetic ions (3d(Fe)) is substantially large ($\sim$8 \AA). In
spite of this, the Curie temperature is very high, exceeding that
of the celebrated manganites (T$_C<$360K). This observation
suggests that the Double Exchange model used to describe the
ferromagnetism in manganites cannot be safely used in the present
case. Sarma et al. \cite{Sarprl} proposed that due to the Fe-Mo
hybridization, the intra-atomic exchange in Mo is much enhanced
thus resulting in a strong antiferromagnetic coupling between Fe
and Mo and thus the leading to an effective more robust Fe-Fe
ferromagnetic ordering. Recently, Fang et al \cite{Fang} have
proposed that ferromagnetism is stabilized by the exchange
splitting of the Mo(4d) orbitals, which lowers the energy of
carriers and promotes a charge transfer from the
$\it{spin}$-$\it{up}$ to the $\it{spin}$-$\it{down}$ subbands.

Based on neutron diffraction \cite{Kobnat,Nak,Morito} and early
M\"{o}ssbauer spectroscopy data \cite{Naka}, ferrimagnetism was
proposed to be originated from the AFM ordering of Fe$^{3+}$
(3d$^5$; t$_{2g}^{3\uparrow}$ e$_{g}^{2\uparrow}$):
Mo$^{5+}$(4d$^1$; t$_{2g}^{1\downarrow}$) configurations, thus
predicting a saturation magnetization of 4$\mu_B$. However, the
experimental values of the saturation moment is commonly found to
be of about 3.1-3.2$\mu_B$ \cite{Kobnat,Naka,Itoh,Tomioka}. So
far, this sensible diminution of the saturation moment has been
attributed to a partial Fe-Mo disorder. From more recent
M\"{o}ssbauer data, however, a state of valence-fluctuation of
Fe$^{2.5+}$ has been proposed by Lind\'{e}n et al. \cite{Linden}
and Balcells et al. \cite{Balc}, and this has been sustained by
Chmaissen et al. who has found $\mu_{Fe}$=4.3-4.4$\mu_B$ from a
simultaneously study based on neutron diffraction and
M\"{o}ssbauer techniques \cite{Chmai}. We recall that, as
indicated in \cite{Balc}, the saturation magnetization values can
not allow discriminating among Fe$^{3+}$: Mo$^{5+}$ or Fe$^{2+}$:
Mo$^{6+}$ electronic configurations. The effective moment in the
paramagnetic phase shall be insensitive to the presence of Fe/Mo
disorder and should, in principle, to allows discrimination
between Fe$^{3+}$:Mo$^{5+}$ or Fe$^{2+}$:Mo$^{6+}$ configurations,
thus it may provide a more robust insight into the electronic
configuration of the Fe/Mo ionic species. Using this approach,
Tovar et al. 18 have recently shown that the magnetic properties
in the paramagnetic regime cannot be understood by considering
only the contribution of localized moments: the effective
paramagnetic moment is found to be smaller that it should be
expected for any electronic configuration Fe (3d$^6$):Mo(4d$^0$)
or Fe(3d$^5$):Mo(4d$^1$) \cite{Martinez}.

In fact, the paramagnetic susceptibility data has been modeled
\cite{Tovar} by assuming that there is an exchange induced spin
polarization of the conduction band being antiferromagnetically
coupled to the localized moments by Hund coupling. Even more, the
strength of the ferromagnetic coupling has been predicted to be
proportional to the density of states at the Fermi level
(D(E$_F$)). These findings may provide microscopic understanding
of the observed augmentation of the Curie temperature upon
electron doping \cite{Navarro} although direct evidence is still
lacking. Similarly, evidence of a half-metallic ferromagnetic
nature of these electron-doped double perovskites is still
missing.

Here, we report on synchrotron radiation photoemission
measurements near the Fermi level of Sr$_{2-x}$La$_x$FeMoO$_6$
with a gradual level of electron doping. As it has been recently
shown \cite{Navarro}, the doping (achieved via partial
substitution of Sr$^{+2}$ by La$^{+ 3}$) promotes a substantial
enhancement of the Curie temperature (with $\Delta$T$_C$ up to
80K). These experimental evidences, sharply contrast with some
recent predictions \cite{Alonso} thus illustrating the complexity
of phase diagram in double perovskites. We will show here, that as
the La substitution progresses, a noticeable enhancement of the
D(E$_F$)-mainly of Mo(4d) parentage is measured. A clear
correlation between D(E$_F$) and T$_C$ is discovered. In addition,
we provide spectroscopic evidence that the electron injection
supplies carriers to the metallic spin-down channel; whilst the
other Fe-3d full-filled (t$_{2g}$$^3$ and e$_g$$^2$) spin-up
insulating channel remains unchanged. These results constitute a
stringent test for proposed models for electronic structure and
ferromagnetism in SFMO and shall provide guidelines for further
progress on tailoring ferromagnetic metals having optimal
properties for spintronics.

\section{Experimental Details}

Photoemission spectroscopy (PES) experiments on
Sr$_{2-x}$La$_x$FeMoO$_6$ ceramic samples have been performed at
the SU8 beamline at LURE using synchrotron radiation light between
14 till 890 eV provided by an insertion device of Super-Aco
storage ring, \cite{Huttel,Davila}. The Fermi level of the oxides
have been determined using Cu as reference. In the range of
explored energies h$\nu$ the energy resolution is of about 50 meV
\cite{Davila,Arranz,Bengio}. The measured intensity has been
normalized with respect to the intensity of the deep O(2p) states.
Due to the polycrystalline nature of the investigated samples, the
results reported here should be considered as angle-integrated
data, because the lack of momentum discrimination in the
reciprocal space. Sample preparation, structural and magnetic
characterization can be found elsewhere
\cite{Balc,Martinez,Navarro}. Here, in order to illustrate the
high quality of the used samples, we only mention that in the
pristine compound ($x$=0), the saturation magnetization is
M$_S$=3.8$\mu_B$ and the antisite concentration (i.e. misplaced
Fe/Mo ions) is $\sim$5\% and the Curie temperature (determined
from extrapolation of the magnetization curves) is of about 430K.
For the rest of samples the corresponding values are T$_C$= 440K,
465K and 480K for $x$=0.2, 0.4 and 0.6 respectively. These T$_C$
values are comparable to those reported in Ref. \cite{Navarro} for
similar samples. The T$_C$ values determined from the Arrot plots
also show the same systematic rise with doping although the
absolute values are somewhat lower; for instance T$_C$=400K and
420K for $x$=0 and 0.4 respectively. Refinement of neutron
diffraction profiles have been used to confirm that, within the
experimental resolution, all samples here reported are oxygen
stoichiometric \cite{Frontera}. Samples have been scratched in
situ in ultra-high vacuum ($<$3 10$^{-11}$ torr) by using a
diamond saw. Surface contamination has been monitored by the
eventual presence of C 1s core level. All measurements have been
carried out at room temperature.

\section{Results}

Figure 1 shows the valence band spectrum of
Sr$_{2-x}$La$_x$FeMoO$_6$ ($x$=0, 0.2, 0.4, and 0.6) samples
collected using h$\nu$=50eV. As it has been reported before for
non-electron doped samples \cite{Kang,Saitoh}, the two major
features, occurring at -8 and -6 eV correspond to O(2p) and
Fe(e$_{t2g}^\uparrow$) states respectively. The metallic edge can
be well recognized. Substantial modifications of the spectra are
observed at or around (0-2eV) the Fermi level upon La
substitution. Within the context of this paper of the greatest
interest is the valence band spectrum close to the E$_F$ as shown
in Figure 2. To elucidate the origin of the observed changes in
the spectra, we have recorded spectra at different photon energies
with the purpose of taking advantage of the dependence of the
photoionization cross sections ($\sigma$) of the different
elements on the photon energy. Detailed inspection of these
spectra immediately reveals a finer structure. At h$\nu$=90 eV
(Fig. 2 (top)), where the cross section of the Fe electronic
states is much higher than those from molybdenum and oxygen
($\sigma$(Fe):$\sigma$(Mo):$\sigma$(O)$\sim$7:0.1:1.5), the
spectra show two main features, one state at the Fermi edge which
changes slightly its intensity as the electron doping progresses
and other state extended from 0.7 eV to 2.5 eV, whose intensity is
unchanged as the La doping increases. In Fig. \ref{Fig_2} (middle)
we show the same spectra recorded at h$\nu$=50eV, where the cross
sections of Fe, Mo and O ions are comparable
($\sigma$(Fe):$\sigma$(Mo):$\sigma$(O)$\sim$ 9:3.5:6). Both
features are present; however their behavior as a function of
doping is different: in this case, the intensity of both states is
clearly rising as the La content increases, thus revealing a
gradual modifications of Mo and O derived states. We include in
Figure 2 (bottom) the theoretical density of states for the parent
compound as recently calculated by Saitoh et al \cite{Saitoh}. In
agreement with computations \cite{Kobnat,Wu,Fang,Saitoh}, we note
that the experimental features observed in the PES can be assigned
to: oxygen O(2p)hybridized spin-up Fe(eg) (-1 to -2.5 eV) states
and spin-down Fe(t$_{2g}$)+Mo(t$_{2g}$) states (0 to -1 eV).

Deconvolution of the experimental data has been done (Fig. 3 )
using a background gaussian function that collects the tail of the
-6 eV peak (Fig. 2) plus three gaussian peaks, labeled C, B and A
convoluted with the Fermi function in Fig. 3. The solid line
through the data shows the quality of fits. This decomposition
allows a clear identification of the states (B at $\sim$-2.0 eV
and C at $\sim$-1.25 eV), corresponding to the spin-up Fe(e$_g$)
doublet band predicted by several theoretical works
\cite{Kobnat,Kobprb,Pickett,Wu,Sarprl,Fang,Saitoh,Kang}. The
expected theoretical band width of 1.5 eV for the spin-up
Fe(e$_g^\uparrow$) states agrees quite well with the states
labeled as B and C. Taking into account the spectral weight
broadening due to the relative high measurement temperature, we
can estimate the photoemission half-gap on the occupied state side
of the majority spin-up Fe(e$_g^\uparrow$) channel to be
0.7$\pm$0.2eV. The peak labeled A, extending up to the Fermi
level, can be associated to spin-down Fe(t$_{2g}^\downarrow$) and
Mo(t$_{2g}^\downarrow$) states. The comparison of the PES spectra
with theoretical predictions \cite{Kobnat,Saitoh} shown in Fig. 2
confirms the peak assignation we have made. From data in Fig. 2,
it is clear that the position of the spin-up Fe(e$_g^\uparrow$)
states (B and C), does not change noticeably with La doping,
indicating that the spin-up (majority) channel gap is not
perturbed by the electron injection all along the investigated
doping regime.

It is important to notice that, in agreement with recent
predictions \cite{Alonso}, the results shown in Fig. 3 indicate
that a simple rigid-band model cannot be used to analyze the data.
In this framework, it shall be expected that a charge transfer or
carrier doping may promote a rigid upward shift of the Fermi level
in order to fill with the extra charge \textit{available
unoccupied bands}. If this were the case the binding energy of
both spin channels should have been increased in the same amount
as the Fermi level may have been shifted. In spite of this, only
the states at the Fermi level (feature A of Fig. 2) associated to
the delocalized spin-down Mo states are modified. The spectral
weight related to those totally polarized states increases and
their binding energy shifts slightly away from the Fermi level. In
contrast, features B and C associated to the spin-up channel are
not affected by the electron doping keeping their binding energy
unchanged for different La content. Consequently, the
photoemission half-gap on the occupied state side of the minority
spin-up Fe(e$_g^\uparrow$)  channel remains invariable ($<$
0.7$\pm$0.2eV) irrespectively on doping.

In order to determine the origin of the progressive rising of the
intensity of peaks A, B and C, we concentrate our analysis on the
PES spectra (Fig.2) obtained at the Mo Cooper minimum (h$\nu$=90
eV), with purpose to discriminate the relative contribution of the
Fe and Mo to the corresponding states. Top part of Fig. 2 reveals
that at h$\nu$=90 eV the t$_{2g}^\downarrow$ states do not change
their intensity upon La doping, thus indicating that the
enhancement observed at h$\nu$=50 eV is due to the some
progressive admixture of Mo (A) and oxygen hybridized orbitals (B
and C), to the bottom of the conduction and top of the valence
band respectively. Of fundamental importance is the observation
that at h$\nu$=50eV (and to lower extent also for h$\nu$=90 eV)
the intensity of the t$_{2g}^\downarrow$ peak (labeled A)
increases and thus the corresponding density of states at the
Fermi level D(E$_F$) also rises. This is a key result that reveals
that La doping and the accompanying electron injection promote an
enhancement of D(E$_F$). The fact that this effect is almost
canceled when photons insensitive to Mo states are used, suggests
that the electron doping supplies charge almost exclusively to
previously unoccupied spin down $\downarrow$ Mo states, although
these states may be strongly hybridized with oxygen and Fe states.

The density of states D(E$_F$) has been evaluated by integrating
the measured PES intensity measured at 50 eV (Fe and Mo sensitive,
Fig. 2(middle) over an energy range of $\pm$100meV around the
Fermi edge inflexion point. As shown in Figure 4 (inset), where we
collect the normalized D(E$_F$) vs La concentration ($x$) (solid
circles), there is a roughly linear enhancement of D(E$_F$) upon
doping. We have tested the robustness of this result by
integrating the PES intensity over other energy ranges
(50-200meV). No significant variation of the D(E$_F$,$x$)
dependence is found. The relevance of this finding can be better
appreciated in Fig.4 (main panel) where we plot the D(E$_F$)
values together with the Curie temperature of each
Sr$_{2-x}$La$_x$FeMoO$_6$ sample. A striking, almost linear,
dependence of T$_C$ on the D(E$_F$) is obtained. This is a
fundamental result that reveals and illustrates the role of the
itinerant carriers on the ferromagnetic coupling in these oxides.
The prediction that T$_C$ may rise when increasing the D(E$_F$)
contained in the formalism of analysis of effective moment and
ferromagnetic coupling recently developed by Tovar et al.
\cite{Tovar} was at the heart of the attempts to rise T$_C$ by
electron doping \cite{Navarro}. It is worth to notice that the
density of states projected on the Fe state, as determined from
the PES obtained at h$\nu$=90 eV (open circles in Fig. 4 (inset)),
does not show any significant variation upon La doping thus
illustrating that doping carriers occupy mainly Mo orbitals.

The data shown here do not provide insight into the microscopic
mechanism for the modification of the density of states, although
it is clearly triggered by the La doping. The difficulty arises
due to the fact that the carrier injection associated to the La
doping is accompanied by a gradual cell expansion and structural
distortion and the enhanced presence of antisites \cite{Navarro}.
These effects are the result of the different sizes of La/Sr ions
and a reduced driving force for Fe/Mo ordering due to the electron
injection. The latter would be consistent with doping electrons
occupying Mo-4d orbitals \cite{Alonso} and thus in agreement with
the present data. The former may reduce the Mo-O-Fe orbital
overlapping thus shrinking the conduction band. However, to what
extent these phenomena contribute to the observed modifications of
the D(E$_F$) is nowadays unknown and further efforts to
discriminate between bond bending, antisites and genuine carrier
doping are required to address this issue.

\section{Conclusions}

In summary, we have provide evidenced that, in double perovskites,
there is a close connection between the D(E$_F$) and the strength
of the ferromagnetic coupling and thus the Curie temperature. This
experimental observation should provide a solid guide for research
of half-metallic ferromagnetic oxides having still higher T$_C$
and consequently opportunities for further developments of
materials for spintronics. We shall mention that recent findings
in ferromagnetic diluted semiconductors \cite{Dietl} also fit in
the framework and methodologies developed here and thus the
present results may have impact in areas of major current
activity. On the other hand, W. Pickett \cite{Pic2} recently
proposed that A$_2$MM'O$_6$ oxides could be ideal candidates for
searching exotic spin-compensated half-metallic antiferromagnetism
and eventually single spin superconductivity. Our observation that
the spin-down carrier density can be adjusted by appropriate
doping may provide an alternative way to reach the required fully
spin compensation.

\begin{acknowledgments}
We thank the AMORE (CEE), LURE, MAT 1999-0984-CO3 and MAT
2002-03431 projects for financial support and the MCYT-LURE for
making available the synchrotron radiation light.
\end{acknowledgments}

%%%%%%%%%%%%%%%%%%%%%%%

%%%%%%%%%%
%%%%%%%%%% FIGURES
%%%%%%%%%%

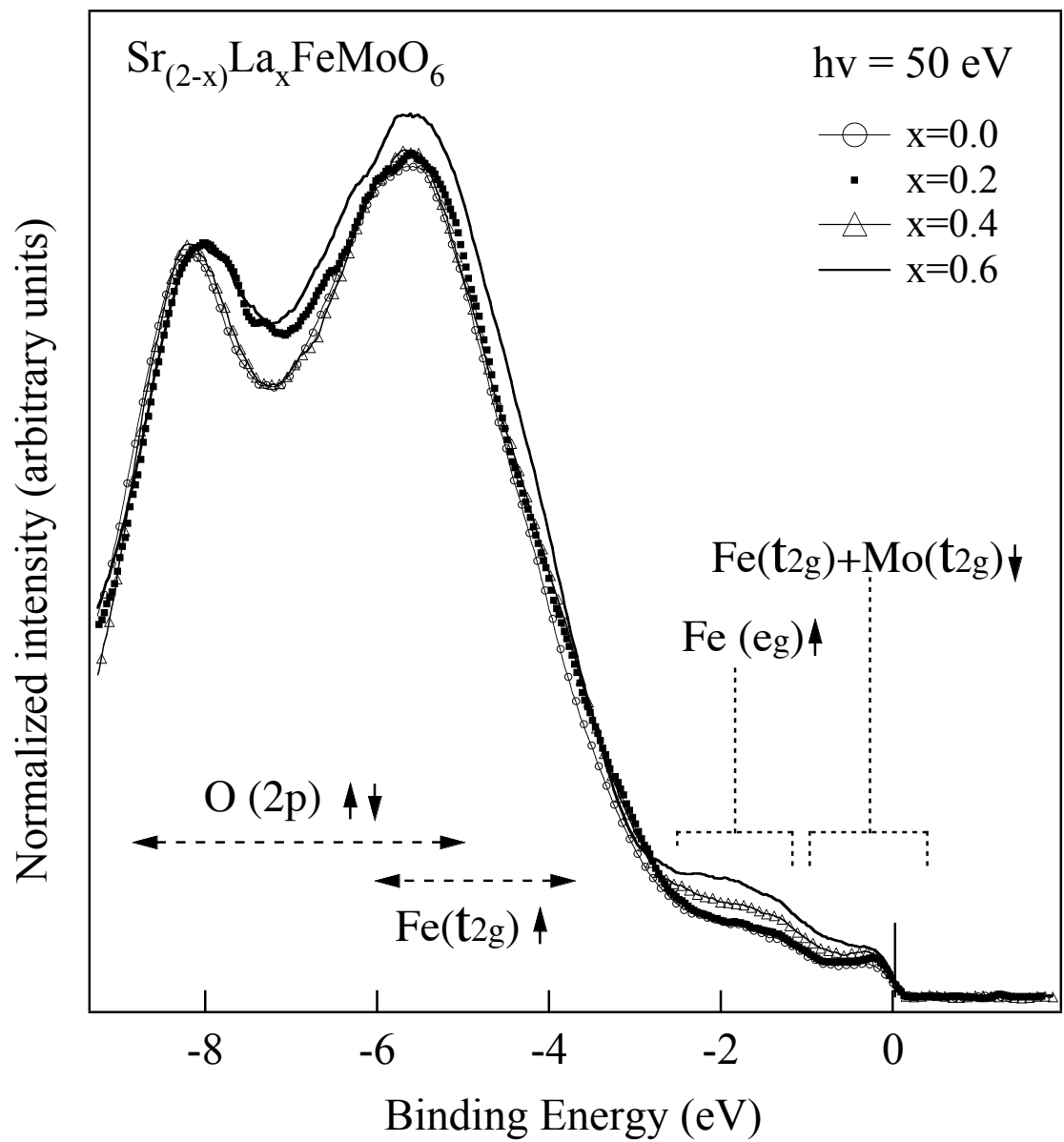
\begin{figure}
\label{Fig_1} \caption{Valence band spectra photoemission spectra
for Sr$_{2-x}$La$_x$FeMoO$_6$ ($x$=0-0.6) recorded at
h$\nu$=50eV.}
\end{figure}

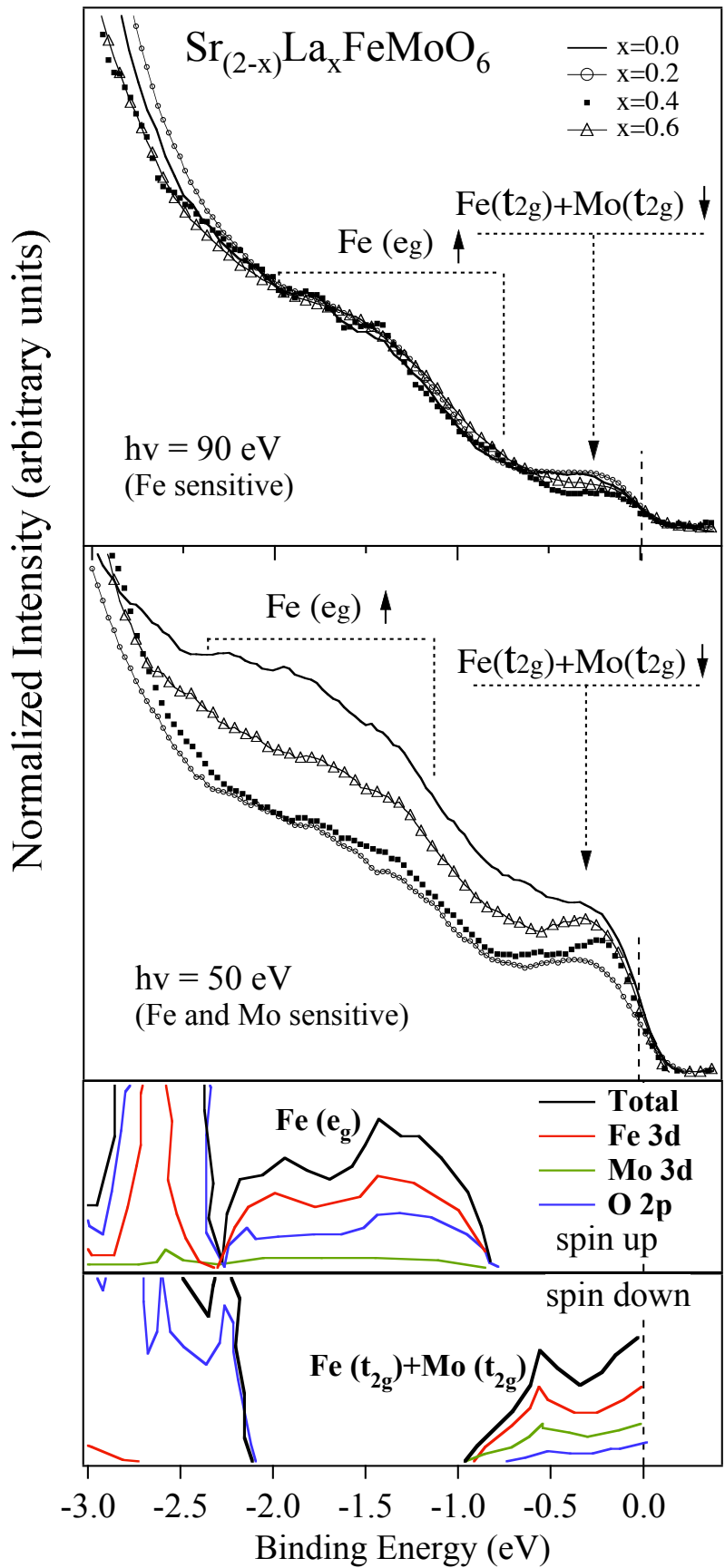
\begin{figure}
\label{Fig_2} \caption{Photoemission spectra near E$_F$ region of
Sr$_{2-x}$La$_x$FeMoO$_6$ ($x$=0-0.6) recorded at h$\nu$=90
eV(top) and h$\nu$=50 eV (middle). In the bottom part, theoretical
calculations adapted from Ref. 29 are included}
\end{figure}

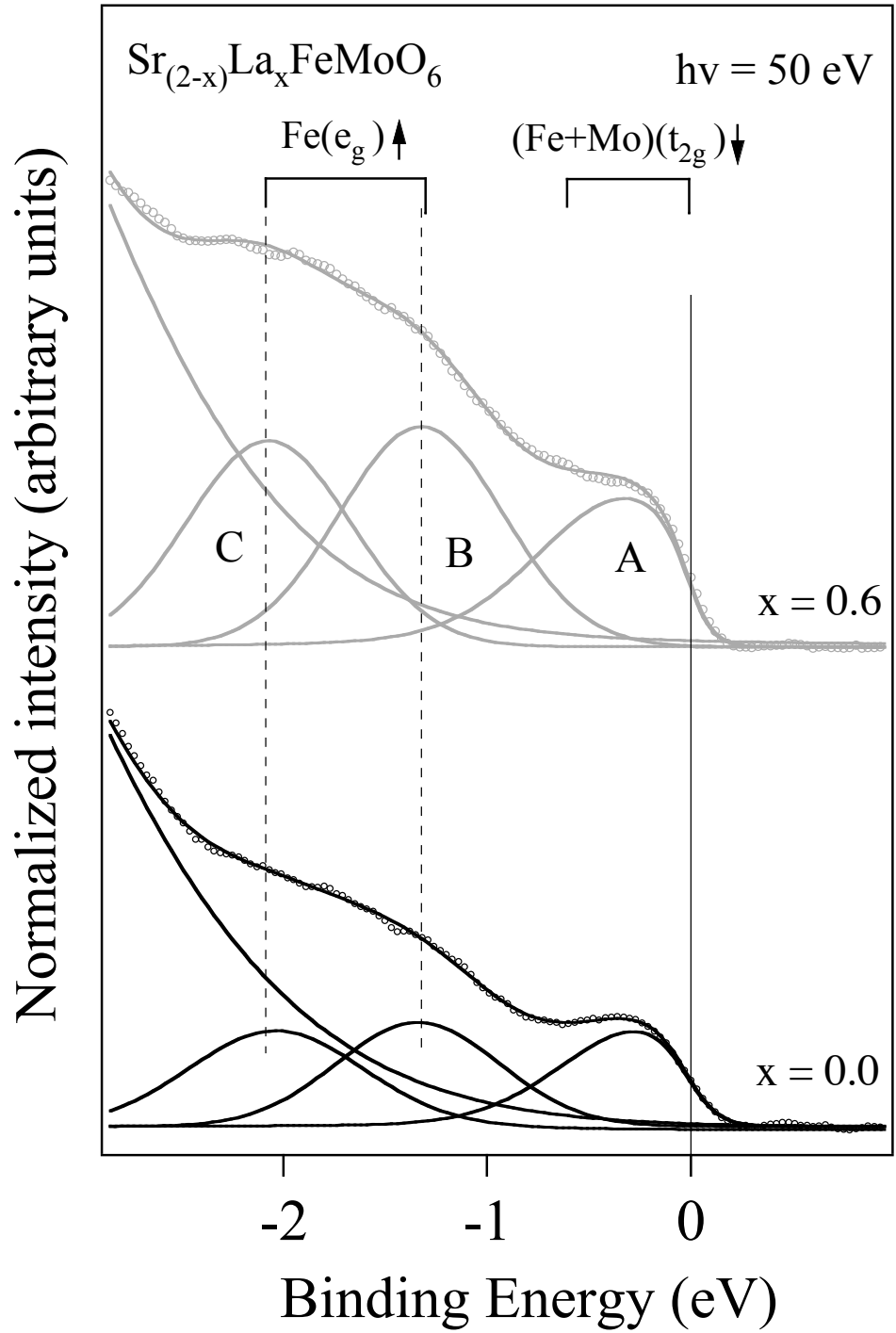
\begin{figure}
\label{Fig_3} \caption{Upper part of the valence band spectra for
Sr$_{2-x}$La$_x$FeMoO$_6$ ($x$=0-0.6), recorded at h$\nu$=50 eV.
Deconvolution of different states are indicated.}
\end{figure}

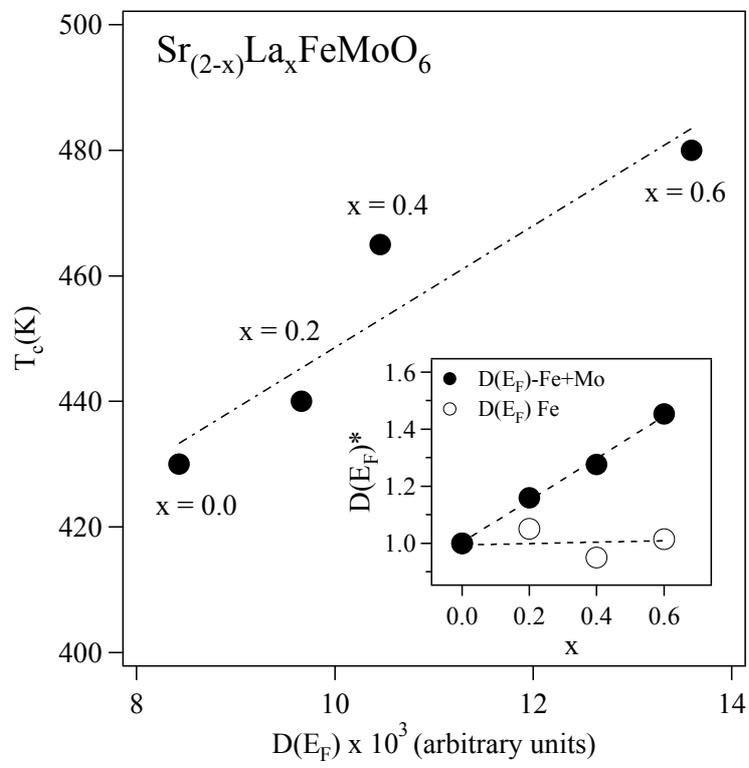
\begin{figure}
\label{Fig_4} \caption{Curie temperature vs the density of states
D(E$_F$,$x$) taken at 50 eV. Inset: Normalized D(E$_F$)=
D(E$_F$,$x$)/D(E$_F$,0)) taken at 50 eV($\bullet$) and 90eV
($\circ$) vs the La contents. }
\end{figure}

\end{document}